\documentclass[12pt, a4paper]{article}

\usepackage{longtable}
\usepackage{bm,a4wide}
\usepackage{ams}
\usepackage{amsmath,amssymb,latexsym,theorem}
\usepackage{bm}
\usepackage{rotating}
\usepackage{graphicx}
\usepackage{hyperref}
\usepackage[dvips]{color}

\begin{document}

\title{$M_F$-dependent Hyperfine Induced Transition Rates in an External Magnetic Field for Be-like $^{47}$Ti$^{18+}$}

\author{\small Jiguang Li $^{1}$~\footnote{Present address: Chimie Quantique et Photophysique, CP 160/09, Universit\'{e} Libre de Bruxelles, Brussels B-1050, Belgium.}, Chenzhong Dong $^{1, 2}$~\footnote{Correspondence author: Dongcz@nwnu.edu.cn}~, Per J\"onsson $^{3}$ and Gediminas Gaigalas $^{4, 5}$ \medskip\\
{\footnotesize $^1$ College of Physics and Electronic Engineering, Northwest Normal University, Lanzhou 730070, China}\\[-0.05cm]
{\scriptsize $^2$ Key Laboratory of Atomic and Molecular Physics \& Functional Materials of Gansu Province, Lanzhou 730070, China}\\[-0.05cm]
{\footnotesize $^3$ Center for Technology Studies, Malm\"o University, Malm\"o S-20506, Sweden}\\[-0.05cm]
{\footnotesize $^4$ Department of Physics, Vilnius Pedagogical University,  Student\c{u} 39, Vilnius LT-08106, Lithuania}\\[-0.05cm]
{\footnotesize $^5$ Institute of Theoretical Physics and Astronomy, A. Gostaut\v{o} 12, Vilnius LT-01108, Lithuania}
}

\date{}

\maketitle

\begin{abstract}
Hyperfine induced $2s2p~^3P_0 \rightarrow 2s^2~^1S_0$ transition
rates in an external magnetic field for Be-like $^{47}$Ti were
calculated based on the multiconfiguration Dirac-Fock method. It was found that the transition
probability is dependent on the magnetic quantum number $M_F$ of the excited state,
even in the weak field. The present investigation clarified that the difference of the hyperfine induced transition rate of Be-like Ti ions
between experiment [Schippers {\sl et al.}, Phys Rev Lett {\bf 98}, (2007) 033001(4)] and theory does not result from the influence of external magnetic field.

\vspace{1cm}

PACS: 31.30.Gs, 32.60.+i

Keywords: Hyperfine induced transition; Zeeman effect; MCDF method.

\end{abstract}

\section{Introduction}
The hyperfine induced transition (HIT) rate of the $2s2p~^3P_0$ level for
Be-like $^{47}$Ti ions has been measured with high accuracy by means of resonant electron-ion recombination in the
heavy-ion storage-ring TSR of the Max-Planck Institute for Nuclear
Physics, Heidelberg, Germany \cite{Schippers}. However, the measured transition rate $A_{HIT}=0.56(3)$ s$^{-1}$ differs from all present
theoretical results $A_{HIT} \thickapprox 0.67$ s$^{-1}$ \cite{Cheng, Andersson, Li} by about 20\%. In the theoretical calculations
the major part of the electron correlation, which always causes the dominant uncertainty, has been taken into account very elaborately. As a result, it is desirable to find out other reasons for the difference.

In this letter, we focus on the influence of the magnetic field present in the heavy-ion storage-ring on the HIT rate. The HIT rate in an external magnetic field depends on the magnetic quantum number $M_F$ of the excited state, even in a relatively weak field. This effect, combined with the non-statistical distribution of the magnetic sublevel population of the excited level, might lead to the difference in transition rate mentioned above.

\section{Theory}
In presence of the magnetic field, the Hamiltonian of an atom with
non-zero nuclear spin $I$ is
\begin{equation}\label{H}
H = H_{fs} + H_{hfs} + H_{m},
\end{equation}
where $H_{fs}$ is the relativistic fine-structure Hamiltonian that
includes the Breit interaction. $H_{hfs}$ is the hyperfine
interaction Hamiltonian, which can be written as a multipole expansion
\begin{equation}
H_{hfs} = \sum_{k \le 1}{\bf T}^{(k)} \cdot {\bf M}^{(k)},
\end{equation}
where ${\bf T}^{(k)}$ and ${\bf M}^{(k)}$ are spherical tensor operator in electronic and nuclear space, respectively \cite{Schwartz}. $H_{m}$ is the interaction Hamiltonian with the external homogeneous magnetic field {\bf B},
\begin{equation}
H_{m} = ({\bf N}^{(1)} + \Delta {\bf N}^{(1)}) \cdot {\bf B},
\end{equation}
where ${\bf N}^{(1)}$ are first-order tensor with the similar form of ${\bf T}^{(1)}$, $\Delta {\bf N}^{(1)}$ is the so called Schwinger QED correction \cite{Cheng2}.

We choose the direction of the magnetic field as the
$z$-direction, and only $M_{F}$ is a good quantum number. The
wavefunction of the atomic system can thus be written as an expansion
\begin{equation}
\label{AW}
|\Upsilon \widetilde{\Gamma} I M_F \rangle = \sum_{\Gamma J F} d_{\Gamma J F} |\Upsilon \Gamma I J F M_F \rangle.
\end{equation}
The total angular momentum $\textbf{F}$ is coupled by the nuclear
$\textbf{I}$ and electronic $\textbf{J}$ angular momentum. The
$\Upsilon$ and $\Gamma$ are the other quantum numbers labeling the
nuclear and electronic states, respectively.

The coefficients $d_{\Gamma J F}$ in Eq. (\ref{AW}) are obtained
through solving the eigenvalue equation using HFSZEEMAN package \cite{Andersson2}
\begin{equation}
{\bf H d} = E \bf{d},
\end{equation}
where ${\bf H}$ is the interaction matrix with elements
\begin{equation}
\label{HME}
H_{\Gamma J F, \Gamma' J' F'} = \langle \Upsilon \Gamma I J F M_F | H_{fs} + H_{hfs} + H_{m}| \Upsilon \Gamma' I J' F' M_F \rangle.
\end{equation}
The readers are referred to Ref. \cite{Cheng2, Andersson2} for
a detailed derivation of the different matrix elements .

For the present problem, the wavefunction of the $^3P_0$ state can be
written
\begin{equation}
\label{WF-1}
|``2s2p~^3P_0 ~ I ~ M_F "  \rangle = d_0 | 2s2p~^3P_0 ~ I ~ F(=I) ~ M_F \rangle + \sum_{S(=1,3); F'} d_{S;F'} |2s2p~^{S}P_1 ~ I ~ F' ~ M_F \rangle.  \end{equation}
The quotation marks in the left-hand wave function emphasize the
fact that the notation is just a label indicating the dominant
character of the eigenvector. Remaining interactions between
$2s2p~^3P_0$ and higher members of the Rydberg series can be
neglected due to large energy separations and comparatively weak
hyperfine couplings \cite{Brage}. Furthermore, those perturbative
states with different total angular momentum $\textbf{F}$ can be
neglected because of relatively weak magnetic interaction. As a
result, Eq. (\ref{WF-1}) is simplified to
\begin{equation}
\label{WF-2}
|``2s2p~^3P_0 ~ I ~ M_F "  \rangle = d_0 | 2s2p~^3P_0 ~ I ~ F(=I) ~ M_F \rangle + \sum_{S=1,3} d_S |2s2p~^{S}P_1 ~ I ~ F(=I) ~ M_F \rangle.
\end{equation}
Similarly, the wavefunction of the ground state is approximatively
written
\begin{equation}
\label{WF-3}
|``2s^2~^1S_0 ~ I ~ M_F "  \rangle = | 2s^2~^1S_0 ~ I ~ F(=I) ~ M_F \rangle,
\end{equation}
where all perturbative states were neglected for the same reasons as mentioned above.

The one-photon $2s2p~^3P_0 \rightarrow 2s^2~^1S_0$ E1 transition
becomes allowed via mixing with the perturbative states of $2s2p~^3P_1$
and $2s2p~^1P_1$ (see Eq. (\ref{WF-2})) induced by both the off-diagonal hyperfine interaction and the
interaction with the magnetic field. The decay rate $a(M^e_F)_{HIT}$ from the excited
state $|``2s2p~^3P_0 ~ I ~ M^e_F "  \rangle$ to the ground state
$|``2s^2~^1S_0 ~ I ~ M^g_F "  \rangle$ in s$^{-1}$ is given
by
\begin{equation}
\label{MHIT-1} a(M^e_F)_{HIT} = \frac{2.02613 \times 10^{18}} {\lambda^3}
\sum_{q} |\langle ``2s^2~^1S_0 ~ I ~ M^g_F " | P^{(1)}_{q} |
``2s2p~^3P_0 ~ I ~ M^e_F " \rangle |^2,
\end{equation}
Substitute Eq. (\ref{WF-2}) and (\ref{WF-3}) into above formula, then
%\begin{scriptsize}
\begin{align}
\label{MHIT-2}
a(M^e_F)_{HIT} &= \frac{2.02613 \times 10^{18}}
{\lambda^3} \sum_{q} |\sum_{S} d_{S} \sqrt{2F^g(=I)+1}\sqrt{2F^e(=I)+1} \nonumber \\
                              & \times \left(\begin{array}{ccc}
                                            F^g(=I)     & 1 & F^e(=I) \\
                                            -M^g_{F(=I)} & q & M^e_{F^e(=I)}
                                            \end{array}\right ) \left \{ \begin{array}{ccc}
                                                                             J^g(=0) & F^g(=I)  & I \\
                                                                             F^e(=I)& J^e(=1) & 1
                                                                         \end{array}\right \} \langle 2s^2~^1S_0 || P^{(1)} || 2s2p~^{S}P_1 \rangle |^2.
\end{align}
%\end{scriptsize}
Applying standard tensor algebra, the Eq. (\ref{MHIT-2}) is further
simplified to
\begin{equation}
\label{MHIT-3}
a(M^e_F)_{HIT} = \frac{2.02613 \times 10^{18}} {3\lambda^3} (2I+1) \sum_{q} | \sum_{S} d_{S} \left ( \begin{array}{ccc}
                                                                                                    I   & 1 & I \\
                                                                                                   -M^g_I & q & M^e_I
                                                                                               \end{array}\right ) \langle 2s^2~^1S_0 || P^{(1)} || 2s2p~^{S}P_1 \rangle |^2,
\end{equation}
where $\lambda$ is the wavelength in {\AA} for the transition and
$\langle 2s^2~^1S_0 || P^{(1)} || 2s2p~^{S}P_1 \rangle$ the reduced electronic transition matrix element in a.u.. 

From the Eq. (\ref{MHIT-3}) we can obtain the Einstein
spontaneous emission transition probability \cite{Cowan}
\begin{align}
\label{TMHIT}
A(M^e_F)_{HIT} &= \sum_{M^g_F} a(M^e_F)_{HIT}  \nonumber \\
         &= \frac{2.02613 \times 10^{18}} {3
\lambda^3} | \sum_{S} d_{S} \langle 2s^2~^1S_0 || P^{(1)} ||
2s2p~^{S}P_1 \rangle|^2.
\end{align}

It should be noticed that in present approximation of weak magnetic field, i.e., neglecting those perturbative states with different total angular quantum number $F$, the formula for the transition rate (see Eq. \ref{TMHIT}) is similar to the one where the transition is induced by only hyperfine interaction \cite{Cheng, Andersson}. However, a significant difference exists in the mixing coefficients $d_{S}$ by virtue of incorporating the magnetic interaction into the Hamiltonian for the present work.

The electronic wavefunctions are computed using the GRASP2K program
package \cite{grasp2K}. Here the wavefunction for a state labeled
$\gamma J$ is approximated by an expansion over $jj$-coupled
configuration state functions (CSFs)
\begin{equation}
|\gamma J \rangle = \sum_i c_i \Phi(\gamma_i J).
\end{equation}
In the multi-configuration self-consistent field (SCF) procedure
both the radial parts of the orbitals and the expansion coefficients
$c_i$ are optimized to self-consistency. In the present work a
Dirac-Coulomb Hamiltonian is used, and the nucleus is described by
an extended Fermi charge distribution \cite{Parpia}. The
multi-configuration SCF calculations are followed by relativistic CI
calculations including Breit interaction and leading QED effects. In
addition, a biorthogonal transformation technique introduced by
Malmqvist \cite{Malmqvist,Biotra} is used to compute reduced transition matrix elements
where the even and odd parity wave functions are built from independently optimized
orbital sets.

\section{Results and discussion}
As a starting point SCF calculations were done for the configurations belonging to the even and odd complex of $n=2$, respectively. Valence correlation was taken into account by including CSFs obtained by single (S) and double (D) excitations from the even and odd reference configurations to active sets of orbitals. The active sets were systematically increased up to $n \le 5$. The SCF calculations were followed by CI calculations in which core-valence and core-core correlations and the Breit interaction and QED effects were incorporated. Based on this correlation model, we calculated the hyperfine induced $2s2p~^3P_0 \rightarrow 2s^2~^1S_0$ E1 transition rate for Be-like $^{47}$Ti ions in absence of the magnetic field to $A_{HIT} = 0.66$ s$^{-1}$, where the experimental wavelength 346.99 \AA ~\cite{NIST} was used to re-scaled the rate.\footnote{The nucleus of $^{47}$Ti has the nuclear spin $I=5/2$, nuclear dipole moment $\mu = -0.78848$ in $\mu_N$ and electric quadrupole moment $Q=0.3$ in barns \cite{Stone}.} The value is in good agreement with the other theoretical results: $A_{HIT} = 0.67$ s$^{-1}$ by Cheng \textit{et al.} \cite{Cheng} and $A_{HIT} = 0.677$ s$^{-1}$ by Andersson \textit{et al.} \cite{Andersson}.

Recent theoretical calculations are all in disagreement with the experimental measurement $A=0.56(3)$ s$^{-1}$ \cite{Schippers} by about 20\%. It is hypothezised that the discrepancy results from the effect of magnetic field present in the storage ring. Actually, the magnetic field effect has already been noticed and been discussed in previous experiment measuring the lifetime of the hyperfine state of metastable level $5d~^4D_{7/2}$ for Xe $^+$ using the ion storage ring CRYRING at the Manne Siegbahn Laboratory (Stockholm) \cite{Mannervik}. Returning to the present problem, experiment was conducted in the heavy-ion storage-ring TSR where the rigidity of the ion beam is given as $B \times \rho = 0.8533$ T \cite{Schippers}, and the bending radius of the storage ring dipole magnets is $\rho = 1.15 $m \cite{Baumann}. As a result, the magnetic field in the experiment has been 0.742 T. Considering the factual experimental environment, we calculated the hyperfine induced $2s2p~^3P_0 \rightarrow 2s^2~^1S_0$ E1 transition rate of Be-like $^{47}$Ti ion in the external magnetic field B=0.5 T, B=0.742 T and B=1 T, respectively.  With assistance of Eq. (\ref{MHIT-3}) and Eq. (\ref{TMHIT}), we obtained the transition rate $a(M^e_F)_{HIT}$ from the excited Zeeman state to the ground Zeeman state, the Einstein transition probability $A(M^e_F)_{HIT}$ of the excited state, and the corresponding lifetime $\tau$.  Computational results are displayed in Table 1. As can be seen from this table, the transition rates $A(M^e_F)_{HIT}$ for each of the individual excited states $``2s2p~^3P_0 ~ I ~ M^e_F "$ are obviously different because the mixing coefficients $d_S$ in Eq. (\ref{TMHIT}) depend on the magnetic quantum number $M^e_F$ of the excited state.

As can be found from Table 1, the lifetime of $^3P_0$ level is still not sensitive to the sublevel specific lifetimes, if the magnetic sublevels are populated statistically (the lifetimes $\tau = \sum_{M^e_F} \tau(M^e_F)/ (2I+1) = 1.52s, 1.52s, 1.53 s$ in the external magnetic field B=0.5T, 0.742T and 1T, respectively). In this case, the zero-field lifetime within the exponential error can be obtained, as made in Ref. \cite{Schippers}, through only a fit of one exponential decay curve instead of 6 exponential decay curves with slightly different decay constants. To the contrary, in the experiment measuring the HIT rate of the $2s2p~^3P_0$ level of the Be-like Ti ion, the level concerned was produced through beam-foil excitation \cite{Schippers2}. As we know, the cross sections with magnetic sublevels for ion-atom collision are different \cite{Mies, Stoehlker}, and the magnetic sublevel population is in general not statistically distributed. Combining this fact with the $M_F$-dependent HIT rate in an external field, the transition probability of $^3P_0$ level cannot be obtained by statistical average over all magnetic sublevel. However, we also noticed that an external magnetic field can lower the transition rate only for those magnetic sublevels with $M_{F} \ge 0$. In other word, only if these specific magnetic sublevels with $M_{F} \ge 0$ were populated, it is possible to explain or decrease the discrepancy between the measured and theoretical HIT rates for Be-like $^{47}$Ti. In fact, such extreme orientation of the stored ions seems improbable by means of beam-foil excitation. Moreover, the experimental heavy-ion storage-ring was only partly covered with dipole magnets (this fraction amounts to 13\%) \cite{Baumann}. It further reduces the influence of magnetic field on the lifetime of level. Therefore, we still cannot clarify the disagreement between experimental measurement and theoretical calculations at present even though the influence of an external magnetic field was taken into account.

\section{Summary}
To sum up, we have calculated the hyperfine induced $2s2s~^3P_0 \rightarrow 2s^2~^1S_0$ E1 transition rate in an external magnetic field for each of the magnetic sub-hyperfine levels of $^{47}$Ti$^{18+}$ ions based on the multiconfiguration Dirac-Fock method. It was found that the transition rate is dependent on the magnetic quantum number $M^e_F$ of the excited state, even in relatively weak magnetic fields. Considering the influence of an external magnetic field, we still did not explain the difference in the HIT rate of Be-like Ti ion between experiment and theory.

\section*{Acknowledgment}
We would like to gratefully thank Prof. Stefan Schippers and Prof. Jianguo Wang for helpful discussions. The referees' very valuable suggestions should be acknowledged. This work is supported by the National Nature Science Foundation of China (Grant No. 10774122, 10876028), the specialized Research Fund for the Doctoral Program of Higher Education of China (Grant No. 20070736001) and the Foundation of Northwest Normal University (NWNU-KJCXGC-03-21). Financial support by the Swedish Research Council is gratefully acknowledged.

\clearpage

\begin{table}
\centering
\begin{footnotesize}
\caption{Hyperfine induced $2s2p~^3P_0 \rightarrow 2s^2~^1S_0$ E1
transition rates in presence of magnetic field B=0.5 T, B=0.742 T and B=1 T for Be-like $^{47}$Ti ion. $a$ represents the
transition probability from the excited state $``2s2p~^3P_0 ~ I ~
M^e_F "$ to the ground state $``2s^2~^1S_0 ~ I ~ M^g_F "$, $A$ is
the Einstein transition probability from the excited state
$``2s2p~^3P_0 ~ I ~ M^e_F "$. $\tau$ is the lifetime of excited state
$``2s2p~^3P_0 ~ I ~ M^e_F "$. The experimental wavelength ($\lambda$) 346.99
\AA \cite{NIST} was used in this calculations, where the influence of hyperfine interaction and magnetic
field was neglected.}
\begin{tabular}{crrccccccccccccccccccc}
\hline
\hline
      &       &    & \multicolumn{3}{c}{B=0.5 T}                 && \multicolumn{3}{c}{B=0.742 T}      && \multicolumn{3}{c}{B=1 T}           \\
\cline{4-6}\cline{8-10}\cline{12-14} M$^e_F$ & M$^g_F$ & $\Delta$ M & $a$ (s$^{-1}$) & $A$ (s$^{-1}$) & $\tau$ (s) && $a$ (s$^{-1}$) & $A$ (s$^{-1}$) & $\tau$ (s) && $a$ (s$^{-1}$) & $A$ (s$^{-1}$) & $\tau$ (s)\\
\hline
5/2  & 5/2  & 0  & 0.44 & 0.61 & 1.64 && 0.42 & 0.59 & 1.71 && 0.40 & 0.56 & 1.78\\
     & 3/2  & -1 & 0.17 &      &      && 0.17 &  &          && 0.16 &  &         \\[0.2cm]
3/2  & 5/2  & 1  & 0.18 & 0.63 & 1.59 && 0.18 & 0.62 & 1.62 && 0.17 & 0.60 & 1.67\\
     & 3/2  & 0  & 0.16 &      &      && 0.16 &  &          && 0.15 &  &         \\
     & 1/2  & -1 & 0.29 &      &      && 0.28 &  &          && 0.27 &  &         \\[0.2cm]
1/2  & 3/2  & 1  & 0.30 & 0.65 & 1.54 && 0.30 & 0.65 & 1.55 && 0.29 & 0.64 & 1.56\\
     & 1/2  & 0  & 0.02 &      &      && 0.02 &  &          && 0.02 &  &         \\
     & -1/2 & -1 & 0.33 &      &      && 0.33 &  &          && 0.33 &  &         \\[0.2cm]
-1/2 & 1/2  & 1  & 0.35 & 0.67 & 1.49 && 0.35 & 0.68 & 1.48 && 0.35 & 0.68 & 1.47\\
     & -1/2 & 0  & 0.02 &      &      && 0.02 &  &          && 0.02 &  &         \\
     & -3/2 & -1 & 0.31 &      &      && 0.31 &  &          && 0.31 &  &         \\[0.2cm]
-3/2 & -1/2 & 1  & 0.32 & 0.69 & 1.44 && 0.32 & 0.71 & 1.41 && 0.33 & 0.73 & 1.38\\
     & -3/2 & 0  & 0.18 &      &      && 0.18 &  &          && 0.19 &  &         \\
     & -5/2 & -1 & 0.20 &      &      && 0.20 &  &          && 0.21 &  &         \\[0.2cm]
-5/2 & -3/2 & 1  & 0.20 & 0.71 & 1.40 && 0.21 & 0.74 & 1.35 && 0.22 & 0.77 & 1.30\\
     & -5/2 & 0  & 0.51 &      &      && 0.53 &  &          && 0.55 &  &         \\[0.2cm]
\hline
\end{tabular}
\end{footnotesize}
\end{table}

\end{document}